# MD simulations and experiments of plasma proteins adsorption to the biodegradable magnesium alloys to facilitate cell response

Xian Wei[a,b*], Jiajia Meng[a*], Sujie Ma[a], Yanchun Li[a], Hong Qing[c], Xubiao Peng[a§], Bo Zhang[a§], and Qing Zhao[a,d§]

[a]*School of Physics, Beijing Institute of Technology, Beijing 100081, China*

[b]*Department of science, Taiyuan Institute of Technology, Taiyuan 030008, China*

[c]*School of Life Science, Beijing Institute of Technology, Beijing 100081, China*

[d]*National Key Laboratory of Science and Technology on Materials under Shock and Impact, Beijing 100081, China*

**ABSTRACT**

Once the magnesium alloy was implanted in the body, it was immediately covered with plasma proteins. The coated alloy surface promoted the adsorption and growth of osteoblasts. Herein, in vitro biological reactions of the ZK60 and AZ31 magnesium alloys were analyzed with and without plasma proteins incubation. The plasma proteins adsorbed on the magnesium alloy were characterized using mass spectrometry (MS). The MS results show that proteins related to bone cells such as fibrinogen, vitronectin, fibronectin, and prothrombin are prone to adsorbed on the surface of the alloys than other proteins. These proteins restrain the degradation of Mg alloys and promote the adsorption and growth of bone cells, which demonstrated by the immersion tests and biocompatibility assays. Furthermore, molecular dynamics simulations were used to analyze the details of the adsorptions of fibrinogen, fibronectin, and prothrombin on ZK60 and AZ31at atomic level. It is revealed that the type of residues adsorbed on the surface of the material has an important effect on protein adsorption.

## 1. Introduction

Magnesium alloy biomaterials exhibit good biocompatibility and degradability, thus, they have

---

[*] Xian Wei and Jiajia Meng contributed equally to this work and should be considered co-first authors.

[§]Corresponding authors: Xubiao Peng, E-mail addresses: xubiaopeng@bit.edu.cn

   Bo Zhang, E-mail addresses: bozhang_quantum@bit.edu.cn

   Qing Zhao, E-mail addresses: qzhaoyuping@bit.edu.cn

unique advantages as implantable materials and are widely used in orthopedics, cardiovascular, and other medical fields [1][2][3]. As an implantable bone substitute, magnesium alloy has an elastic modulus of 41–45 GPa similar to that of human bone, which can effectively avoid stress-shielding effects [4][5]. Magnesium alloy also has excellent mechanical properties and light-weight ratio, providing good mechanical support for implantation; in addition, it does not require a second operation for its removal after the tissue heals because it degrades completely, thus reducing pain [6][7][8]. Moreover, the magnesium ions released during the degradation of magnesium alloys are essential trace elements for the human body, participating in important physiological activities [9][10]. However, the success of biomaterial implantation is largely determined by the biological reaction that occurs on its surface after implantation [11]. After the magnesium alloy is implanted in the body, its first contact is with plasma proteins then the adsorbed proteins will selectively bind to specific cells to activate cell response factors, thereby triggering a series of biological reactions, such as blood coagulation and inflammation [12][13][14]. For bone implant materials, thrombus formed by blood coagulation reaction caused by implant has a positive effect on bone growth [15]. Therefore, the adsorption of plasma proteins plays an important role in the biocompatibility of magnesium alloys and can affect the adsorption and growth of bone cells to a certain extent [16][17].

High-throughput proteomics has been used to explore the interactions between proteins and materials [18]. Mass spectrometry (MS) can be used to identify complex protein mixtures adsorbed on the surface of the materials and obtain the types and relative abundance of individual proteins [19][20]. Abdallah et al. [21] used MS to study the key basal lamina proteins (laminin, nidogen-1) adsorbed on the surface of special biological materials (such as aminated PMMA and aminated PDLLA) and found that this material can promote the adsorption of specific proteins and improve its interaction with epithelial cells. Milleret et al. [22] identified and quantified more than 100 plasma proteins adsorbed on materials by MS and analyzed the main proteins. The results showed that the surface cobalt–chromium oxide alloy adsorbs early plasma proteins causing blood activation reaction afterward, which is of positive significance. Emphasizing the importance of understanding and controlling protein-material interactions can improve and guide the performance of medical implants.

Molecular dynamics (MD) simulations are used to theoretically study the interactions between proteins and materials at atomic level [23][24]. Köhler et al. [25] used MD to study the initial stage of adsorption of fibrinogen on the surface of mica and graphite, revealing that fibrinogen exhibits a weak adsorption force on mica, and frequent adsorption and desorption phenomena occur. Wu et al. [26] used MD to investigate the adsorption of proteins on the surface of pure titanium dioxide and titanium dioxide contaminated with formic acid and discovered that BSA can be stably adsorbed on a pure surface. However, when the surface is contaminated with formic acid, because only the C–H bond is exposed, the polarity of the surface is reduced and the interaction between protein and surface is significantly reduced.

Because the density function theory (DFT) can be used to obtain the charge distribution of different materials, the L–J parameter and charge parameter are added to the force field parameters [27][28][29]. Therefore, MD can be used to simulate the interaction between magnesium alloy and protein. Wang et al. [30] used MD to simulate the interaction between magnesium alloy doped with different alloying elements and fibrinogen, revealing that electronegativity has an important influence on protein adsorption.

After implantation of the magnesium alloy in the body, the adsorption of protein on surfaces will cause a series of cell reactions and affects the degradation of Mg alloys. In this study, fluorescence microscopy and MTT experiments were performed to evaluate a series of effects of the magnesium alloy surface adsorbed by plasma proteins on MG63 cells. Subsequently, the relative amount of proteins adsorbed on two AZ31 and ZK60 magnesium alloys surfaces was identified. The effect of the represented proteins on degradation of ZK60 and AZ31 Mg alloys are investigated using immersion tests. In the end, the MD simulations are performed to study the interactions of several important proteins selected from the MS results with alloys ZK60 and AZ31, revealing the internal mechanism of the different protein adsorption phenomena.

## 2. Experimental and simulation methods

### 2.1 Sample preparation

ZK60 and AZ31 magnesium alloys, whose chemical compositions were determined by inductively coupled plasma optical emission spectrometry (ICP-OES, 725-ES, Agilent, USA), are listed in Table 1. The magnesium alloys were cut into 10 mm × 10 mm × 2 mm chips and ground with a series of SiC papers up to 2000 grit. The samples were ultrasonically cleaned in deionized water and ethanol solution for 5 min, respectively, and then dried in air.

### 2.2. Adsorption of proteins

All the Mg alloy surfaces were washed with phosphate-buffered saline (PBS) solution and immediately transferred into 5-mL human plasma samples (Beijing Beina Chuanglian Biotechnology Institute), which were incubated in a water bath for 30 min at 37°C. After incubation with blood plasma, the samples were rinsed 3 times with PBS and used for either the determination of total protein adsorption by the bicinchoninic acid (BCA) method or identification of adsorbed proteins (after digestion and elution) by MS.

### 2.2.1 BCA test

BCA protein assay kit (Beijing Solarbio Science & Technology Co. Ltd, China) was used to analyze the concentration of protein adsorption [31]. Briefly, 150-µL PBS was added to the samples, which were shaken for 1 min and left in the dark at room temperature (RT) (approximately 25°C) for 30 min. Then, 200-µL BCA-working reagent was added to each sample and place at 37°C for 30 min. After incubation, 100-µL supernatant of each specimen was transferred into a well of a 96 well-plate.

The protein concentration was measured through the optical density at 562 nm using a spectrophotometer. The amount of protein adsorption on the samples (A, µg/cm$^2$) was calculated using the following Eq. (1). Three parallel samples under were each condition.

$$A = \frac{C \times V}{S} \quad (1)$$

where $V$ is the volume of plasma used (mL) and $S$ is the total surface area of each sample (cm$^2$).

### 2.2.2 Protein digestion and peptide elution from the ZK60 and AZ31 surfaces

The surfaces were incubated with 6-M urea in 0.1-M Tris-HCl (pH 8.0) buffer. Dithiothreitol (DTT) was added to reach a final concentration of 10 mM. The sample was then shaken for 1 min on a vortex oscillator and incubated for 1 h at 37°C in a water bath. After cooling at RT, iodoacetamide was added (to achieve the final concentration of 20 mM), and the samples were left in the dark at RT for 30 min. Next, 50 mM NH$_4$HCO$_3$ solution was added to the sample to achieve the urea concentration <1 M. The proteins were digested overnight at 37°C with trypsin (Promega, Madison, WI), in a ratio of 1:50 trypsin:protein. To stop trypsin activity, the solution was acidified with 0.5%–1% formic acid (FA). Eluted peptides were collected in fresh Eppendorf tubes.

Peptide solutions were then desalted using Ziptip (Millipore, Switzerland) protocol; briefly, the ZipTips were equilibrated by aspirating and dispensing 10-µL 2% acetonitrile (ACN) and 0.1% FA twice, followed by two rinsing steps with 10-µL double-distilled water (ddH$_2$O). Thus, the samples were loaded onto the ZipTips by repeated aspiration and dispensing for 10 cycles. Next, the ZipTips with bound samples were washed by aspirating 10-μL 2% ACN and 0.1% FA and dispensing five times. Samples were then eluted from the ZipTips with 10-μL 50% ACN and 0.1% FA of aspirating the elution buffer and dispensed in a fresh Eppendorf tube. Then, the peptide solution was dried in a rotary vacuum dryer and stored at −80°C.

### 2.2.3 MS

Samples were dissolved in 0.1% FA and analyzed in an Orbitrap Fusion mass spectrometer (Thermo Fisher Scientific, US). The liquid phase system, EASY-NLC, was used for separation. The peptide sample was loaded on a low-pH reversed-phase C18 capillary chromatography (150 μm × 150 mm, 1.9 μm, Thermo Inc.). The solvent composition of buffer A was 0.1% FA–water solution and buffer B was 0.1% FA–ACN solution. The effective elution gradient was 6%–35%, the total elution time was 60 min, and the flow rate was 0.5 μL/min. Liquid-phase conditions are shown in Table 2.

The mass spectrum was acquired in the high-sensitivity mode of the Orbitrap mass analyzer. Each full scan was a high-speed signal-dependent scan, and the scan time was 60 min. The first-level full scan resolution was 60,000, the scan range 300–1400 m/z, the AGC target for 5e5, the maximum injection time of 50 ms, and the collision energy of 32%. Secondary scanning resolution was 15,000, AGC target 5e4, maximum injection time of 45 ms, charge state screening (containing +2 to +6 charge precursors), dynamic elimination 50 s; the exclusion window was set to 10 ppm.

Qualitative (or relative quantitative) analysis of proteins was performed based on a technical platform of EASY-NLC 1200 ultrahigh performance liquid-phase mass spectrometer combined with Proteome Discovery 2.3 data processing software. The precursor ion mass tolerance was 10 ppm and the fragment ion tolerance was 0.02 Da using the protein database of the corresponding species. The false discovery rate (FDR) of peptides and proteins was <1.0%, and at least one specific peptide was identified for each protein.

### 2.3 Immersion tests

To investigate the effect of the proteins on degradation of ZK60 and AZ31 Mg alloys, two represented proteins (fibrinogen (Fg) and fibronectin (Fn)) were selected to fabricate the media for degradation tests. The prepared media were as follows: 1) NS (normal saline solution), 2) Fg-NS (20 μg/mL fibrinogen (Solarbio, Beijing) in normal saline), 3) Fn-NS (20 μg/mL fibronectin (Solarbio, Beijing) in normal saline). 2.5 mL of media were added into the 24-well plates, respectively. Subsequently, the 24-well plates were incubated under cell culture conditions (37 °C, 5% $CO_2$). After cultured for 20 min, 2 h, and 6 h, The extracts in the 24-well plates were collected in the centrifuge tube and diluted with deionized water to 10 mL. The released metallic ion concentrations were measured through the inductively coupled plasma mass spectrometry (ICP-MS, 7800, Agilent, USA). Three parallel specimens were set for each test. The degradation rates of the samples were calculated by the contents of the metallic ions released in the media [33] as follows:

$$DR = \frac{(C_{test} - C_{initial})V}{At} \qquad (2)$$

where DR is the degradation rate in µg/cm² per hour. $C_{test}$ is the concentration of the metal ions released in the media. $C_{origin}$ is the average concentration of metal ions in the initial media. V is volume of the tested solution. A denotes the surface area of the samples, and t is the immersion time.

## 2.4 Biocompatibility of ZK60 and AZ31 surfaces

MG63 is a satisfactory model to study human bone cells, exhibiting various characteristics of osteoblasts [32]. MG63 cells (American Type Culture Collection, USA) were cultured in Dulbecco's Modified Eagle's Medium (DMEM, Gibco) along with 10% fetal bovine serum (FBS) at 37°C in a humidified atmosphere of 5% $CO_2$, achieving a final concentration of $1.2 \times 10^5$ cells/mL. The alloy samples were sterilized using a UV light for 30 min and rinsed with PBS 3 times for subsequent plasma protein incubation and final MG63 cell adsorption.

### 2.4.1 Incubation of MG63 cells adsorbed on the plasma protein surface

MG63 cell solution (1 mL) was added to the ZK60 and AZ31 surfaces, which were previously incubated with blood plasma. After 0.5, 2, and 6 h incubation at 37°C under static conditions, the test surfaces were rinsed 3 times each with PBS followed by fixation with 4% PFA for 15 min. Thus, 2% Triton X-100 was used to permeate for 10 min. Subsequently, osteoblasts-exposed surfaces were stained with a 1:100 Hoechst 33342 solution in PBS for 5 min (Beijing Solarbio Science & Technology Co. Ltd, China). The sample was analyzed using a fluorescence microscope (Nikon A1R HD25 High-Resolution Confocal Microscope, Nikon). Five images of each sample were obtained for quantification. The number of nuclei was measured using the ImageJ software.

### 2.4.2 Cell viability

MTT assay was used to measure the viability rate of cells *in vitro*. The extracts were obtained by immersing the alloys with or without plasma protein adsorption on the surface in a cell-free culture solution with an extraction rate of 1.25 cm$^2$/mL, incubated at 37°C in 5% $CO_2$ for 3 days. Thus, the supernatant fluid was withdrawn, filtered, and stored at 4°C for the subsequent test. MG63 cells were seeded in 96-well plates with a density of 1000 per well for cell attachment. After culturing for 24 h, the medium was changed to alloy extracts, and 10% FBS was added simultaneously. The control group was incubated in DMED with 10% FBS. After 1, 3, and 5 days of culture, 10-μL MTT was added to each well. After 4 h of cocultivation with MTT, 150-μL dimethyl sulfoxide was added to dissolve the formed formazan crystals. At a wavelength of 490 nm, the adsorption peak of each well was measured using a microplate reader (Cytation, Bio-The, USA). The cell viability was calculated as follows:

$$Viability = \frac{OD_{sample}}{OD_{control}} \times 100\% \tag{3}$$

### 2.5. Simulation methods

### 2.5.1 Alloy model construction

Materials Studio was used to build alloy model. AZ31 and ZK60 alloys were constructed by replacing the magnesium atoms in the Mg (0001) surface containing two layers of atoms with 3% aluminum atoms and 5.5% zinc atoms, respectively. The established unit cell was separated using a 20 Å vacuum layer in the *z*-direction to avoid artificial influences in the simulation box. AZ31 and ZK60 contain 3136 and 3600 atoms, respectively, as the slab models. DFT simulation with CASTEP Module was used to determine the atomic charge of AZ31 and ZK60 alloys [34][35]. The exchange-correlation of electrons was described using the generalized gradient approximation (GGA) in the Perdew–Burke–Ernzerh (PBE) [36]. The cutoff energy of the plane-wave basis was set to 550 eV. The convergence threshold for optimized configuration was set to <0.05 eV/Å in force on each atom. The Brillouin zone was sampled by a Gamma-centered mesh with a 3 × 3 × 1 k-point grid.

### 2.5.2 MD simulations on protein-alloy interactions

MD simulations were performed using the GROMACS 5.5.1[37] package to investigate the intrinsic mechanism of adsorption behavior of several important plasma proteins identified by MS on ZK60 and AZ31 alloy slab. The CHARMM36m forcefield[38] was used for the proteins, and the TIP3P model[39] was used for water molecules. The initial structure of the protein in the simulation was properly orientated so that the surface of the protein with more negatively charged residues (ASP, GLU) facing the slab, and the initial distance between proteins and surfaces of the slab was ~5 Å to ensure that the occurrence of spontaneous adsorption. The supercell of ZK60 and AZ31 alloys was frozen during the simulations. The parameters of Lennard–Jones (L–J) potential and charge distribution for AZ31 and ZK60 surfaces were obtained from the literature [40] and DFT method, respectively. The parameters were then transformed to apply to CHARMM36m forcefield using the potential energy of the non-bond interaction

$$U(r_{ij}) = 4\epsilon_{ij}\left[\left(\frac{\sigma_{ij}}{r_{ij}}\right)^{12} - \left(\frac{\sigma_{ij}}{r_{ij}}\right)^{6}\right] + \frac{q_i q_j}{r_{ij}}, \tag{4}$$

where $\epsilon_{ij} = \sqrt{\epsilon_i \epsilon_j}$ and $\sigma_{ij} = (\sigma_i + \sigma_j)/2$ according to the Lorentz–Berthelot combination rule, with $\epsilon_i$ being the well depth of L-J interaction potential and $\sigma_i$ being the particle size of the *i*-th atom . The cutoff distances of the electrostatic interactions and van der Waals (vdW) interactions were both set to 12 Å. The Particle–Mesh–Ewald (PME) method was used for the long-range electrostatic interactions. In the simulation, covalent bonds connected with hydrogen atoms were restricted, the temperature was set to 310 K, the reference pressure was 1 bar, and the timestep was 2fs. The simulation time of each molecule was 10 ns. The whole system for simulation was in a water boxes with size 146.4 Å × 146.4 Å × 160.0 Å.

## 3. Results

### 3.1 Degradation behavior

To investigate the effect of the plasma proteins on the degradation of ZK60 and AZ31 Mg alloys, two representative proteins solutions (Fg and Fn) are selected for test . Fig. 1shows the degradation rates of ZK60 and AZ31 Mg alloys in different media after immersion for 20 min, 2 h, and 6 h, where the degradation rate is calculated based on Mg ion content released in solution. After immersion for 20 min, the degradation rates of the AZ31 alloy decrease significantly in the media with proteins, especially in Fg-NS solution. In contrast, the degradation rates of the ZK60 alloy increase with the addition of protein. With the extension of immersion time, the degradation rate of the samples differs significantly. After incubation for 2 h, the AZ31 alloy shows slower degradation rate in Fg-NS solution compared with that in Fn-NS solution. However, the opposite trend is observed in ZK60 alloys. Both AZ31 and ZK60 alloys have slower degradation rate in the media with proteins, which indicates that the adsorption of proteins played a positive role in inhibiting the degradation of the alloys. As degradation intensifies after immersion for 6h, the degradation of the AZ31 alloy is further inhibited both in Fg-NS and Fn-NS solutions (Fig.1c). However, as for ZK60 alloy, the degradation rate increases in the media with proteins. The significant elevation for ZK60 alloy in Fn-NS solution can be related to the rupture of the adsorbed protein film [41].

## 3.2 Plasma proteins mediate cell adsorption

In general, the biomaterial first interacts with the plasma proteins in the blood after being implanted in the human body, which leads to the adsorption of the plasma proteins on the surface of the implants. Thus, to study the effect of protein adsorption on the cellular response, cell adsorption assay and MTT assay were performed. Fig. 2 shows the fluorescent images of the cell adsorption for the ZK60 and AZ31 surfaces with and without plasma protein incubation. As shown in Fig. 2, a larger number of cells adhered to the surfaces of the pretreated samples compared with that of untreated samples. This implies that the formed plasma protein layer promotes the MG63 cell adsorption. In the absence of the protein incubation, the AZ31 alloy surface exhibited better adhesion than that of ZK60 alloy (Fig. 2e, 2g, 2i, 2k). This can be due to the corrosion of the ZK60 alloy surface in the presence of excess hydrogen gas, which hinders the adhesion of the cells. The ZK60 alloy with an excess of zinc forms a high quantity of precipitates, leading to severe galvanic corrosion compared to the low content of Al in the AZ31 alloy [42]. As the incubation increased, the adsorbed cells increased considerably for both the untreated and pretreated samples because the sample surfaces are adhered to the proteins present in the culture medium, contributing to enhanced cell adsorption. The quantification of cell nuclei (Fig. 2m) confirmed the trend observed for cell adsorption in the fluorescent images. The pretreatment with plasma proteins improves the adsorption of MG63 cells on the alloy surface, particularly in the ZK60 alloy.

Fig. 3a displays the *in vitro* viability of MG63 cells cultured in the extracts of the ZK60 and AZ31 alloys with and without plasma proteins incubation. Based on the ISO 10993-5 [43], when the cell viability exceeds 80%, the biological material does not possess any toxic effect, otherwise, it is toxic. As shown in Fig. 3a, visible cells are observed for both untreated and pretreated samples, indicating superior biocompatibility. After culturing for 3 days, the cell viability is reduced to 80% in the ZK60 extract, while increased considerably in the pretreated ZK60 extract. This is because the protein layer formed on the surface inhibits the degradation of the pretreated samples during the extract process, which is confirmed by the decreased concentrations of $Mg^{2+}$, $Zn^{2+}$, and $Zr^{2+}$ for the pretreated ZK60 alloy in Fig. 3b. According to the work of Wang et al., 15 mM (360 μg/mL) of $Mg^{2+}$ are not toxic to MC3T3-E1 cells[44]. Although $Mg^{2+}$ in the pretreated AZ31 extract are higher than those in the untreated one, the concentration of $Mg^{2+}$ (136.8 μg/mL) released in pretreated AZ31 extract is still too low to cause toxicity (Fig. 3b). Therefore, both the untreated and pretreated samples exhibit extremely high viability after incubation for 3 days, suggesting outstanding biocompatibility. The cell viability cultured by the pretreated AZ31 extract is slightly lower than that of the untreated AZ31, while it shows the opposite trend for the ZK60 case. This implies that the positive effect of the pretreated by plasma proteins on the Zn-containing sample is more significant compared to that of the Al-containing samples, as the incubation proceeds.

**3.3 Adhesion of plasma proteins on the biomaterials**

The plasma proteins exhibit different effects on distinct Mg alloys in altering cell adsorption and toxicity, indicating the important role of protein adhesion to the cells. To specifically understand the mechanism of protein adhesion, the BCA and MS methods were used to analyze the plasma protein layer qualitatively and quantitatively on the alloy surfaces. Fig. 4a shows the total protein adsorption on the surfaces of ZK60 and AZ31 Mg alloys measured by BCA after incubation for 0.5 h. No significant difference in the protein adsorption between the ZK60 and AZ31 alloys was observed. However, the protein types adsorbed on the surfaces of the two alloys were distinct, as demonstrated by the composition of the proteins with high content on the alloy surfaces in Fig. 4b. Some proteins (e.g., complement C3, fibrinogen, α2-macroglobulin, apolipoprotein E and A-I, plasminogen, kininogen) in the plasma exhibited higher levels compared to those adsorbed on the ZK60 and AZ31 alloys. However, the amounts of the proteins vitronectin, fibronectin, and prothrombin adsorbed on the surfaces of the two alloys are higher than those in the plasma, indicating that these proteins possess stronger adsorption capacity on the surfaces of the Mg alloys. The amounts of fibrinogen and vitronectin proteins are higher on the surface of AZ31 alloy than ZK60 alloy, indicating that the AZ31 alloy has a stronger attraction to these proteins. In contrast, more peptide fragments of the prothrombin were observed on the ZK60 alloy surface than those of the AZ31 sample.

**3.4 Simulation on the adhesion of protein**

It is known that the proteins fibrinogen, fibronectin, and prothrombin have positive effects on bone growth. Moreover, the fibronectin and prothrombin proteins have relatively higher level adsorptions on the alloy surfaces compared with the plasma solution. Therefore, these proteins were selected for the theoretically investigation on the mechanism of the protein adsorption on the ZK60 and AZ31 alloys by the MD simulations. The crystal structures of the fibrinogen, fibronectin, and prothrombin (PDBID: 1fid, 1ttg, 3k65) [17][45][46] were obtained from the RCSB as the initial structures for simulation.

These proteins exhibited a net charge of −5e, 0e, and 6e in neutral pH solution, respectively. The initial structures of the proteins on the ZK60 and AZ31 alloy surfaces are shown in Fig. 5, and the parameters of the forcefield for ZK60 and AZ31 Mg alloys are listed in Table 3.

When the protein is interacting with the material, its conformation is actively adjusted until it is adsorbed on the material in a stable conformation. The RMSD of a protein is used to measure whether the structure of the protein is stable. Fig. 6 shows the RMSD of the three proteins on ZK60 and AZ31 alloys, respectively. In the beginning, under the combined action of electrostatic force and vdW force, the RMSD changed sharply. However, the RMSD gradually converged to a constant value of 2/3.5, 3/3.2, and 2/2.8 Å for ZK60/AZ31-1fid, ZK60/AZ31-1ttg, and ZK60/AZ31-3k65, respectively, indicating that the protein gradually adsorbed on the material within 10 ns and reached a relatively stable state.

Figs. 7–8 show the minimal distance and interaction energy between the proteins and alloy surfaces, respectively. The minimal distance is defined as the minimum of the distances between each heavy atom of the protein and the alloy surface. As shown in Fig. 7a, the fibrinogen is originally distant from the ZK60 alloy surface and then it is adsorbed after approximately 1 ns. Nevertheless, it can be directly adsorbed to the surface of AZ31 alloy in a very short time. The fibronectin adhered on the ZK60 Mg alloy surface in the initial stage, and then gradually moved away from the surface after 2 ns, implying that it no longer interacts with the surface over time (Fig. 7b). However, the opposite trend is observed on the AZ31 Mg alloy, in which the fibronectin is adsorbed on the surface after 4 ns and stayed steadily afterwards. The prothrombin is stably adsorbed on both ZK60 and AZ31 alloys surfaces (Fig. 7c).

The interaction energy between the protein and material is an important criterion for detecting whether the protein is adsorbed. A lower interaction energy usually indicates better adsorption of the protein. Fig. 8(a, d) shows a positive total interaction energy with the repulsive electrostatic interaction being dominant, indicating that the protein fibrinogen is less adsorbable on both alloys. The total interaction energy in AZ31 is smaller than ZK60, indicating the adsorption of fibrinogen is relatively more adsorbable on AZ31, which is consistent with the experimental observation in Fig 4(b). When fibronectin is adsorbed on the ZK60 alloy (Fig. 8b), the electrostatic interaction and vdW interaction have similar contributions in the first 2 ns and the interaction energy slightly negative, showing that fibronectin can be better adsorbed on both alloys. However, the adsorption on ZK60 seems unstable in our simulation and the fibronectin flied away from the surface of ZK60 after 2ns. In contrast, the adsorption of fibronectin on AZ31 is very stable after 4ns, with the vdW force being the main driving force for adsorption (Fig. 8e) . When prothrombin is adsorbed on ZK60 and AZ31, Fig. 8(c, f), the vdW plays a leading role in the adsorption process, making the total interaction energy the lowest, which indicates that prothrombin is the best adsorbed on the surface of alloys among the three proteins. In addition, the total interaction energy of prothrombin on ZK60 is clearly smaller than that on AZ31, indicating prothrombin has better adsorption on ZK60 than AZ31. Such observations are again consistent with the results of the mass spectrum displayed in Fig. 4b.

Fig. 9 shows the overall conformations of fibrinogen, fibronectin, and prothrombin when they are adsorbed on ZK60 and AZ31. Fig. 9 demonstrates that the flexible random coils can be better adsorbed on the surface of the substrate compared with the secondary structures like α − helix and β − sheet. It can be explained by considering the fact that the flexible random coil can easily change its local conformation to be better adsorbed on the alloy surface when contacting with the substrate. To further analyze the adsorption details of the three proteins adsorbed on both alloys, the type and number of residues contacting with the alloy surface (within 3.5 Å from the surface) were identified and listed in Table 4. Four out of five residues of fibrinogen adsorbed on the surface of ZK60 and AZ31 are in common, among which there are two hydrophobic residues known as LEU and PHE. Since the numbers of the charged/polar and hydrophobic contacting residues are comparable, both the electrostatic and vdW interactions contribute to the adsorption of fibrinogen on alloys. For fibronectin, the residues adsorbed on the ZK60 alloy are VAL (hydrophobic residue) and ARG (charged residue), while the residues adsorbed on AZ31 are the hydrophobic residues VAL and ALA. According to the types of the contacting residues, the adsorption behavior of fibronectin on ZK60 alloy is due to both the electrostatic interaction and the vdW interaction, while the adsorption on AZ31 alloy is mainly caused by the vdW interaction. For the third protein prothrombin, the contacting residues on the alloys include three hydrophobic residues (PHE, ILE, LEU for ZK60 and PHE, ILE, GLN for AZ31); Hence, the vdW interaction is the main driving force for the adsorption of prothrombin on the two alloys. It is important to compare the number of adsorbed residues for the same protein on different alloys. The larger the number of adsorbed residues, the stronger the interaction between the protein and substrate. For fibrinogen, both ZK60 and AZ31 have five residues adsorbed on the surface during most of the simulation time. However, we notice that after 9 ns there is only one residue can be adsorbed on the surface of ZK60 since the protein fibrinogen floated out of the substrate as shown in the subset of Fig. 7a. In contrast, on AZ31 fibrinogen always have five residues in contact with alloy. Hence, the adsorption of fibrinogen on ZK60 is less stable than on AZ31. Therefore, it is consistent with the higher abundance of fibrinogen adsorbed on the AZ31 alloy in the mass spectrum of Fig. 4b. For Fibronectin, there are more adsorbed residues on AZ31 (3 residues) than ZK60 (2 residues), and the total interaction energy was also lower on AZ31, indicating that the adsorption was better on AZ31, which is again consistent with the MS experimental results showing that the abundance of fibronectin adsorbed on AZ31 was higher than that on ZK60. Similarly, the number of residues adsorbed by prothrombin on the surface of ZK60 alloy (5 residues) was higher than that of AZ31 (4 residues), and the interaction energy was lower than that on the surface of AZ31, indicating that the protein prothrombin adsorbed better on the surface of ZK60. This is consistent with the result that the prothrombin adsorption abundance on the ZK60 surface is larger than that on the AZ31 surface obtained by the MS experiment. Consequently, the type and number of residues adsorbed on the surface of the material significantly affect the vdW and electrostatic interaction forces, which in turn affect the protein adsorption on the material, and ultimately affects the growth and proliferation of cells.

## 4. Discussion

Mg dissolves quickly accompanied with the formation of hydrogen when the samples are immersed in the media. A corrosion product layer composed of $Mg(OH)_2$ is formed on the surfaces of AZ31 and ZK60 Mg alloys. The insoluble $Mg(OH)_2$ layer initially acts as a protective barrier to the samples. The oxidation reduction reactions are as follows:

$$Mg \rightarrow Mg^{2+} + 2e^- \tag{5}$$

$$H_2O + 2e^- \rightarrow H_2 + 2OH^- \tag{6}$$

$$Mg + 2H_2O \rightarrow Mg(OH)_2 + H_2 \tag{7}$$

As the degradation continues, more and more corrosion products are deposited on the surface. Then, the $Mg(OH)_2$ layer is degraded in the solution with aggressive $Cl^-$ ions. The corresponding reaction is as follows:

$$Mg(OH)_2 + 2Cl^- \rightarrow MgCl_2 + 2OH^- \tag{8}$$

The degradation rates of both AZ31 and ZK60 Mg alloys decrease in the solution containing proteins, which is due to the adsorption of proteins (as shown in Fig. 1). However, as the reaction continues, the adsorption of protein on the alloy surface becomes weak, resulting in an increased degradation rate of the alloy. This phenomenon is evident in ZK60 alloys.

After implantation in the human body, the biological implant first contacts the plasma proteins in the blood, resulting in protein adhesion. Then, the implant interacts with the cells, leading to cell adhesion, proliferation, and migration. Therefore, two adhesion steps (proteins adhesion and cell adhesion) are used to evaluate the effects of proteins on the adhesion and viability of MG63 cells. In this study, the tested samples were incubated in the plasma solution for 30 min to activate the biological responses of the whole blood [47][48]. When MG63 cells are seeded on ZK60 and AZ31 surfaces without any proteins adhered, the cell adhesion and viability were not elevated. Instead, if the samples were incubated with proteins before culturing with MG63 cells, cell adhesion and viability were significantly improved. Previous studies have also shown that proteins have a vital control effect on cells, for example, the presence of plasma proteins can regulate the adsorption of platelets and neutrophils [48][49].

Protein adhesion affects cell adsorption in many ways, such as the total amount, the composition of the adsorbed protein layer, and the conformation of adsorbed proteins. To elucidate the key factors of protein layers on the behavior of MG63 cells, the total amount of proteins on the surface of ZK60 and AZ31 were measured, and the results show that the total amount of protein adsorbed on the surfaces of the two alloys is similar. Therefore, the composition of the protein adhered on the two alloys is identified by MS, which is suitable for identifying different types of proteins in the protein mixture [50]. Different types of proteins promote cell adhesion in different ways [51].

Evidently, both ZK60 and AZ31 Mg alloy surfaces adsorbed a large amount of albumin, haptoglobin, and immunoglobulin on the surface (in the Supplementary Material) owing to the high levels of these proteins in the plasma. However, the composition of the protein layer is different on different surfaces, which indicates that different proteins exhibit distinct adsorption capacities on the Mg alloy surfaces. As shown in Fig. 4b, in addition to the high abundance of proteins in the plasma, the amount of fibrinogen, vitronectin, and prothrombin adsorbed on the two alloys are higher than the remaining proteins. Although the content of fibronectin in the plasma is lower, the amount of fibronectin adsorbed on the alloy is higher than that in the plasma, suggesting that fibronectin is more likely to adhere to the two magnesium alloys. Fibrinogen can specifically bind to receptors on the cell surface to promote cell adsorption and growth [52][53]. Furthermore, fibrinogen is considered a key factor in the coagulation reaction, which can promote the adsorption of platelets [54], transform into a fibrin network through thrombin [55], promote coagulation reaction, and form a thrombus. Vitronectin and fibronectin are glycoproteins, which play an important role in cell activation during the initial stage of cell adsorption [56][57]. These two proteins can promote the reorganization of the actin microfilaments and the creation of new focal contacts and can jointly promote the adsorption and spreading of cells on the material [58]. Moreover, fibronectin and vitronectin bind to the $\alpha_5\beta_1$ and $\alpha_v\beta_3$ integrins on the surface of osteoblasts through the RGD (Arg-Gly-Asp) region, respectively, to promote bone differentiation [59][60]. The amino acid sequence RGD is a domain present in each protein of fibrinogen, vitronectin, and fibronectin, which can promote the adhesion and growth of bone cells [4]. Therefore, Figs. 2–3 shows that the pre-adsorbed plasma protein samples promote the adhesion and growth of MG63 cells. Additionally, the adhesion level of fibrinogen and vitronectin on the surface of AZ31 is higher than that of ZK60, which leads to better adhesion of bone cells to the surface of AZ31 alloy. The adhesion amount of prothrombin on the surface of ZK60 is higher than that of AZ31 (Fig. 4), enhancing the possibility of triggering the coagulation reaction. Prothrombin is involved in the internal pathway of blood coagulation [61]. Kininogen of high molecular weight (HMWK), prekallikrein, and factor XII contact the surface (negatively charged) to initiate activation of prothrombin. After a series of reactions, prothrombin is decomposed into coagulation. Thrombin can activate fibrinogen to form a fibrin network and participate in thrombus formation [13]. Although for cardiovascular implants, thrombus formation has serious consequences for the success of implantation, for bone implant materials, higher thrombus formation can better promote osseointegration and growth [63]. Therefore, ZK60 also exhibits excellent properties as a bone implant material.

The adsorption of proteins on materials is a complex process involving many factors. This study focused on the theoretical investigation of the key factors involved in protein adsorption to explain the causes for the phenomenon described above based on MD simulations. The simulation results are consistent with the MS results. Because the relative abundance of fibrinogen adsorbed on the alloy is lower than that in the plasma, the adhesion of fibrinogen on the two alloys is not satisfactory. Although the abundance of fibronectin and prothrombin in the plasma is not elevated, the relative abundance adsorbed on the two alloys is relatively high, indicating that the two proteins are more likely to be adsorbed on the two alloys ZK60 and AZ31. This phenomenon can be explained by the interaction energy between the protein and material [64]. Studies have shown that the vdW interactions can play a major role in the process of protein adsorption to the materials [65]. Additionally, electrostatic interactions are also decisive in the protein adsorption to the materials [66]. Fig. 8 shows that both the vdW and electrostatic interactions have different effects on the protein adsorption to the alloy surfaces owing to the presence of fewer charges on the surface of real alloys. For systems with enhanced adsorption, the vdW plays a leading role in the protein adsorption process. For the adsorption of fibronectin on AZ31 and prothrombin on the two Mg alloys, the absolute value of vdW energy is higher than the electrostatic interaction energy; thus, the vdW energy is the main driving force for the adsorption of fibronectin and prothrombin. Essentially, the interaction between the type of adsorbed residues and the material surface determines the type of interaction energy. As listed in Table 4, the fibronectin adsorbed on AZ31 and prothrombin adsorbed on the two Mg alloys are mainly hydrophobic residues, i.e., VAL, PHE, ILE, LEU, and GLN. Therefore, the vdW energy is greater than the electrostatic energy. For systems with poor adsorption, the electrostatic interaction energy is greater than the vdW energy; thus, the electrostatic interaction plays a major role. The fibrinogen adsorption on the two materials is driven by electrostatic interactions. The main residues adsorbed on the alloy surface are charged residues, i.e., GLU, HIS, LEU, and PHE. The interaction of these charged residues with the substrate is greater than that of the hydrophobic residues with the substrate, resulting in higher electrostatic interaction energy than vdW interaction energy. For the adsorption of fibronectin on the ZK60 alloy, the adsorbed residues include both the charged residue ARG6 and hydrophobic residue VAL1. Therefore, both vdW and electrostatic interaction are essential in the adsorption process, but the vdW is higher than the electrostatic interaction energy, resulting in a relatively satisfactory adsorption capacity.

## 5. Conclusions

In summary, the adsorption of plasma proteins on the surface of the materials can effectively promote the adsorption and growth of bone cells. Moreover, the type and content of the proteins adsorbed on different materials are distinct. The proteins, such as fibrinogen, vitronectin, fibronectin, and prothrombin, tend to be adsorbed on the surfaces of the AZ31 and ZK60 magnesium alloys, which are beneficial to delay the degradation rate and facilitate bone growth. Vitronectin and fibronectin promote cell growth by binding the integrins of the bone cells through the RGD region. Fibrinogen and prothrombin can promote the formation of thrombus and bone growth. MD simulations were used to evaluate the internal mechanism of alloys attracting fibrinogen, vitronectin, prothrombin, and other bone growth-promoting proteins. The interaction energy between the protein and material surface is the main driving force for adsorption. The type of residues adsorbed on the alloy surface determines if the electrostatic interaction or vdW interaction plays a major role. The adsorption of fibronectin and prothrombin on the two alloys is relatively good, which is mainly due to that the hydrophobic residues are adsorbed on the surface and the vdW energy is the main driving force for adsorption. The fibrinogen adsorption on the two alloys is not satisfactory, mostly resulted from the mixture adsorption of charged residues and hydrophobic residues and the synergic action of the electrostatic energy and vdW energy. Therefore, MD simulations can be used as an important approach to predict the adsorption capacity of proteins on different alloy surfaces.

## CRediT authorship contribution statement

**Xian Wei**: Methodology, Writing-Original draft preparation. **Jiajia Meng**: Conducting the experiments, Writing-Reviewing and Editing. **Sujie Ma**: Data analysis and processing. **Yanchun Li**: Simulation analysis. **Hong Qing**: Supervision, Resources. **Xubiao Peng**: Supervision, Formal analysis, Software. **Bo Zhang**: Supervision, Validation. **Qing Zhao**: Conceptualization, Supervision, Resources.

## Declaration of Competing Interest

The authors declare that they have no known competing financial interests or personal relationships that could have appeared to influence the work reported in this paper.

## Acknowledgments

This work was supported by the National Science Foundation (NSF) of China (grant No. 11675014), and the Ministry of Science and Technology of China (2013YQ030595-3). Additional support was provided by Fundamental Research Program of Shanxi Province in China (20210302124264).

**Table 1. Chemical composition of AZ31 and ZK60 alloys.**

| Element | Mg | Al | Zn | Mn | Ca | Na | Si | P | S | Fe | K | Pb | Zr |
|---|---|---|---|---|---|---|---|---|---|---|---|---|---|
| AZ31(wt.%) | Bal | 3.0549 | 1.0047 | 0.224 | 0.0212 | 0.0167 | 0.011 | 0.006 | 0.0055 | 0.0043 | 0.0035 | 0.0022 | — |
| ZK60(wt.%) | Bal | 0.02 | 5.5279 | 0.0247 | 0.0805 | 0.0153 | 0.0327 | 0.0059 | — | 0.039 | 0.004 | — | 0.5847 |

**Table 2. Low-pH reversed-phase chromatography gradient.**

| Time(min) | A% | B% | Flow rate ( μL/min) |
|---|---|---|---|
| 0 | 94 | 6 | 500 |
| 2:00 | 90 | 10 | 500 |
| 40:00 | 78 | 22 | 500 |
| 48:00 | 65 | 35 | 500 |
| 55:00 | 10 | 90 | 500 |
| 60:00 | 10 | 90 | 500 |

**Table 3. Parameters for the ZK60 and AZ31 forcefields.**

| Atom | $q$(e) (ZK60/AZ31) | $\epsilon$ (kJ/mol) | $\sigma$ (Å) |
|---|---|---|---|
| Mg | 0.0013/0.0010 | 0.06270 | 0.211 |
| Zn | -0.0637 | 1.04600 | 0.194 |
| Al | -0.0310 | 0.87802 | 0.145 |

**Table 4. Type and quantity of residues adsorbed on ZK60 or AZ31.**

| Surface and protein | Adsorption Residue | Number of Adsorption Residue |
|---|---|---|
| ZK60-1fid | GLU256 LEU259 PHE160 PRO156 SER157 | 5 |
| AZ31-1fid | PHE160 LEU259 PRO156 HIS257 SER 157 | 5 |
| ZK60-1ttg | VAL1 ARG6 | 2 |
| AZ31-1ttg | VAL1 ALA83 PRO82 | 3 |
| ZK60-3k65 | PHE234 ILE39 LEU35 SER 37 SER25 | 5 |
| AZ31-3k65 | PHE234 ILE39 ARG165 GLN173 | 4 |

**Figure captions**

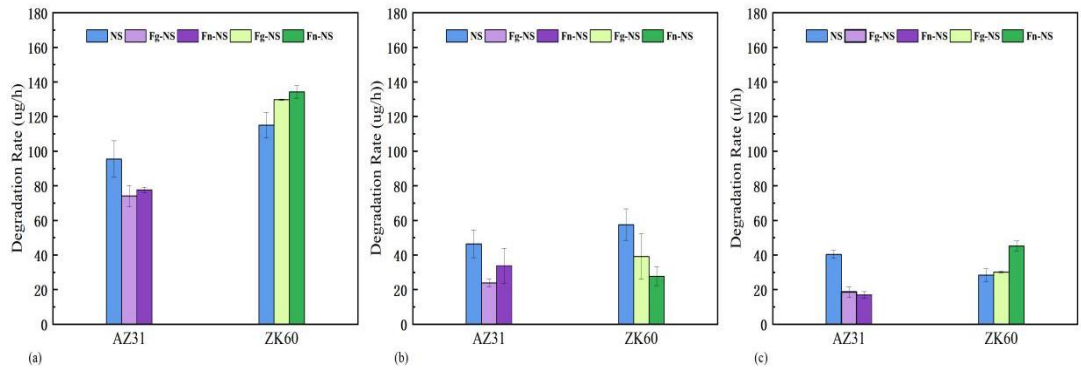

Fig. 1 The degradation rates of ZK60 and AZ31 Mg alloys calculated based on Mg ion content released in different media after immersion for 20 min, 2 h, and 6 h, respectively.

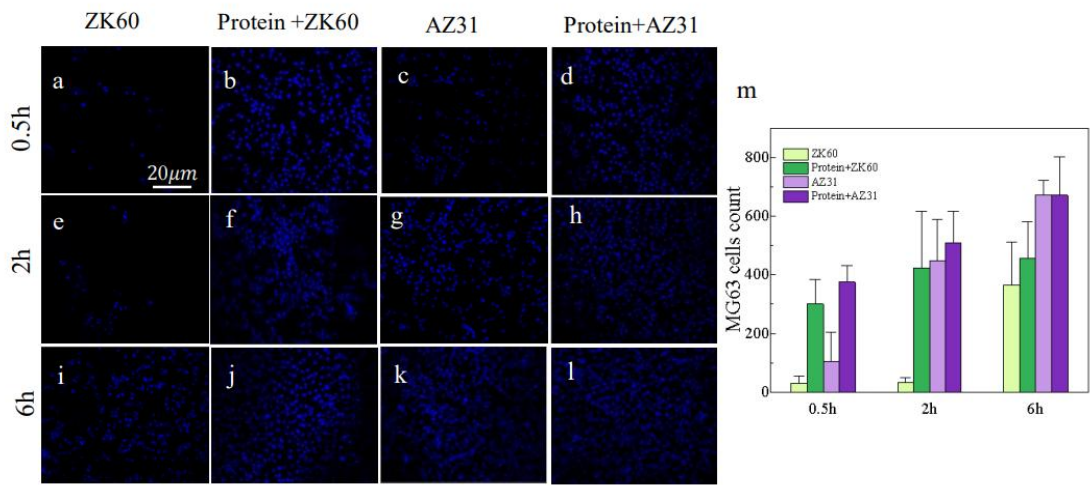

Fig. 2 Fluorescent images of ZK60 with proteins (b, f, and j) and without proteins (a, e, and i), and AZ31 with protein (d, h, and l) and without protein (c, g, and k) surfaces, showing the nuclei of adherent MG63 (blue dots). Cells were incubated in different surfaces in distinct times: 0.5 h (a–d); 2h (e–h); 6 h (i–l); (m) Quantification of the nuclei counts. The plot shows the means and standard deviations of triplicates.

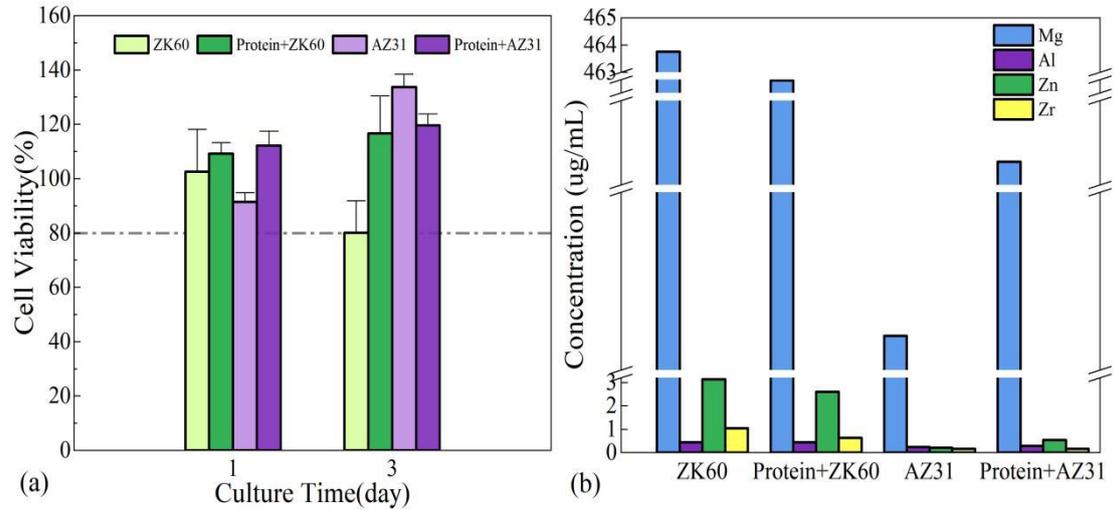

Fig. 3 a) *In vitro* viability of MG63 cells cultured in the extracts of ZK60 and AZ31 with and without plasma proteins for 1 and 3 days b) Concentrations of Mg, Al, Zn, and Zr ions in the extracts of untreated and treated by the plasma proteins for the ZK60 and AZ31 after incubation for 3 days.

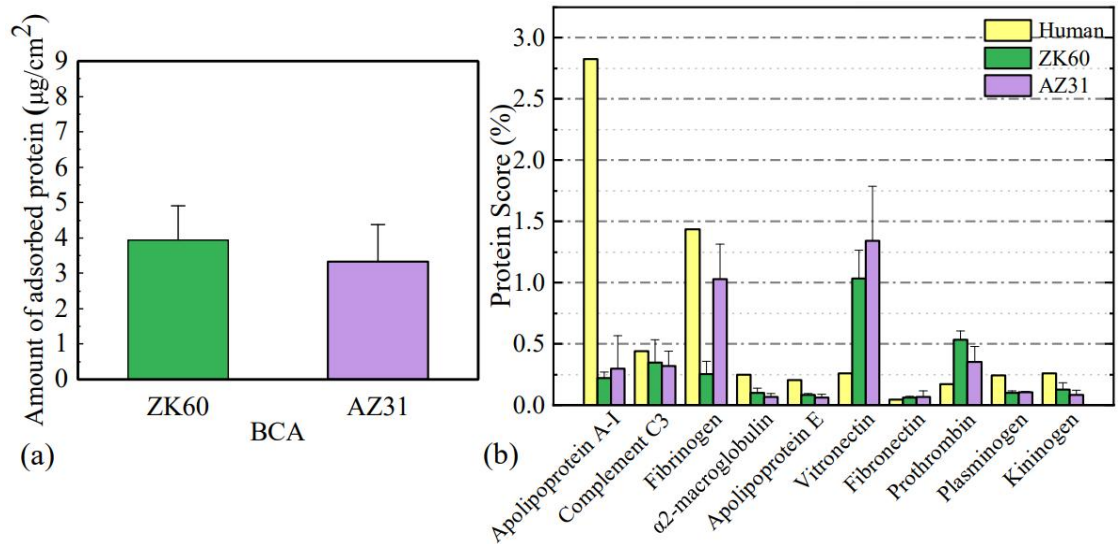

Fig. 4 a) Total protein adsorption on the different materials measured by BCA after incubation for 0.5 h. b) Composition of the proteins with high content on the alloy surfaces determined by MS. The remaining proteins detected, which were not discussed in this study, are displayed in S1 and S2.

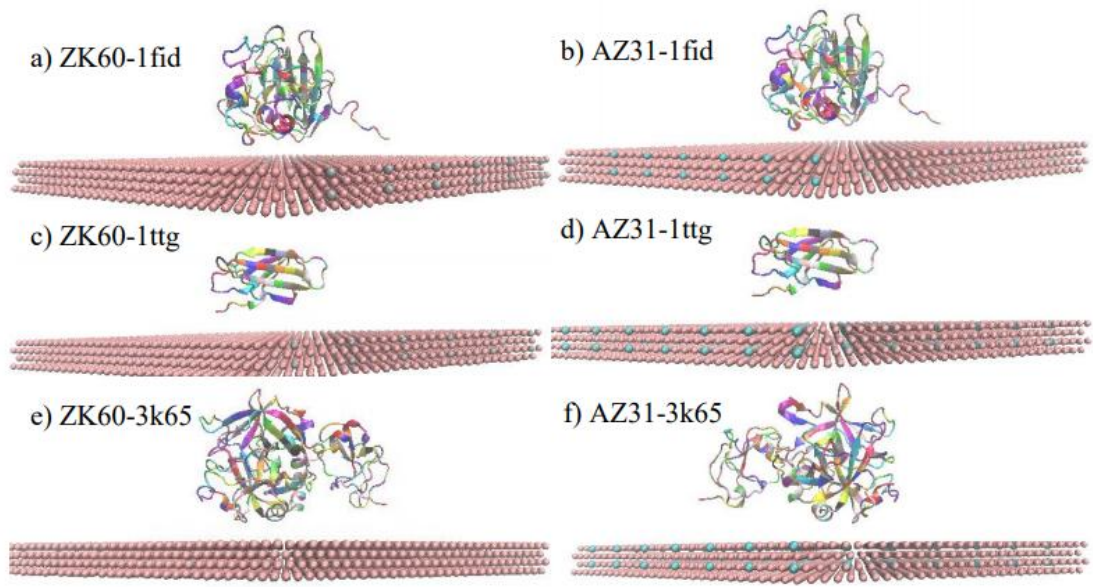

Fig. 5 a–f) Initial conformations of fibrinogen, fibronectin, and prothrombin on the surface of ZK60 or AZ31(ZK60-1fid, AZ31-1fid, ZK60-1ttg, AZ31-1ttg, ZK60-3k65, AZ31-3k65) respectively.

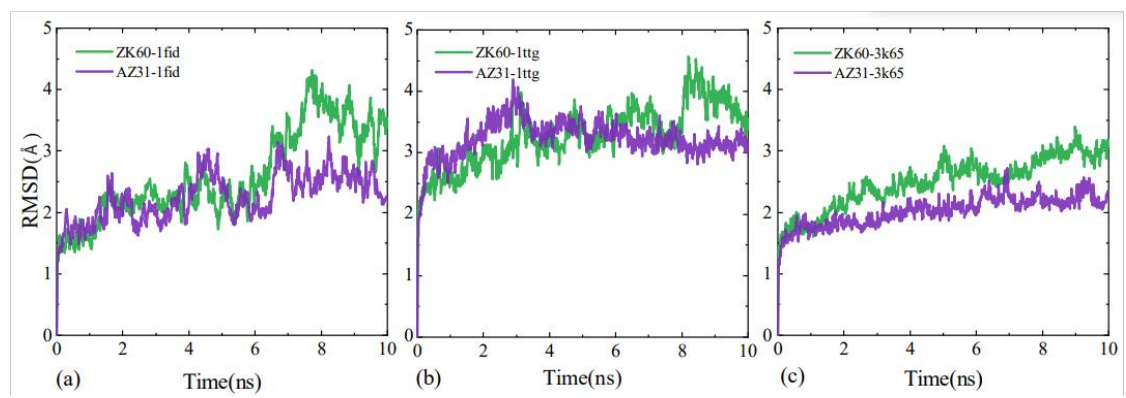

Fig. 6 a–c) RMSD of fibrinogen, fibronectin, and prothrombin during MD simulation on the surface of ZK60 or AZ31, respectively.

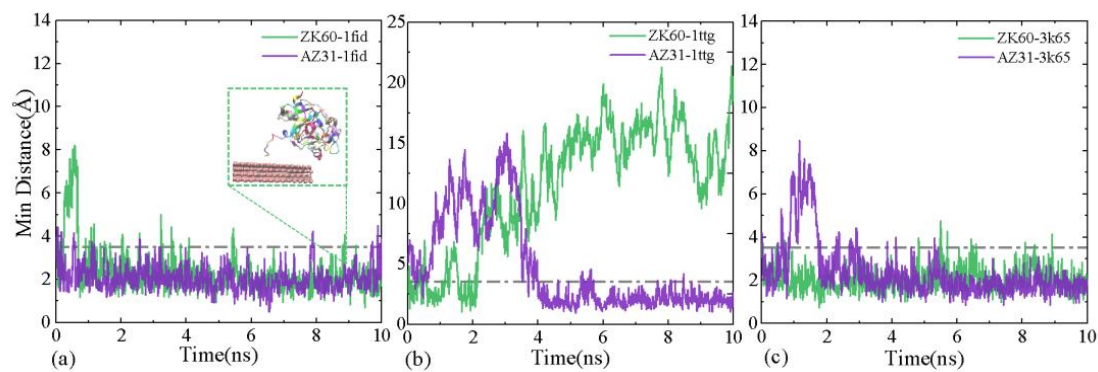

Fig. 7 a–c) Minimum distance between the three proteins (fibrinogen, fibronectin, and prothrombin) and the surface of ZK60 or AZ31, respectively.

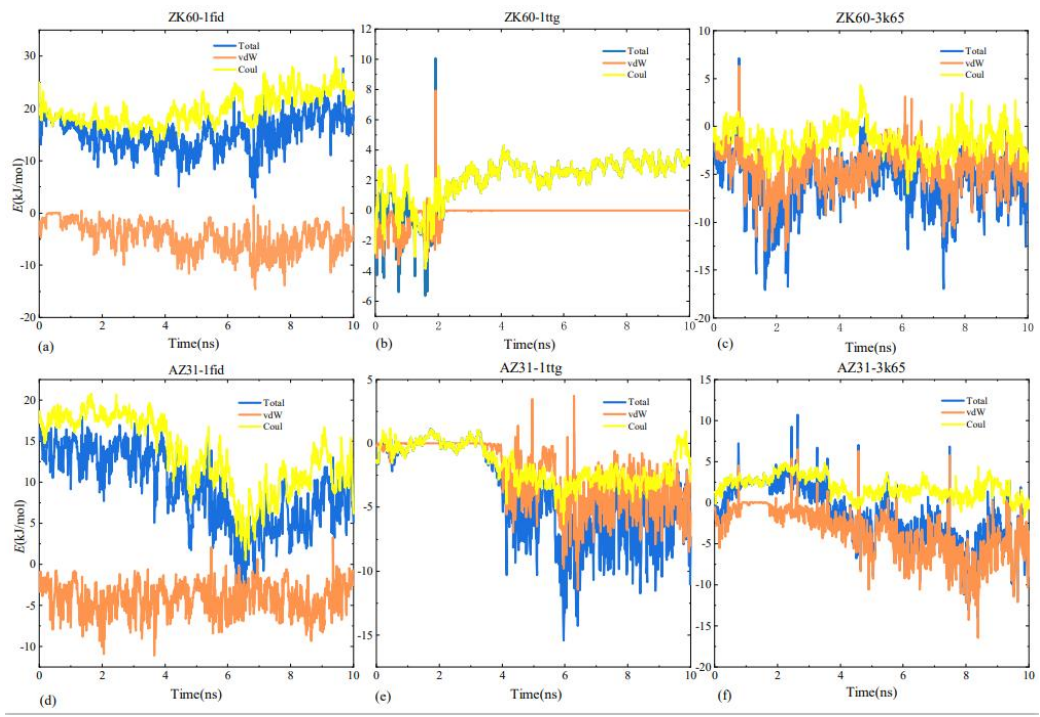

Fig. 8 a–f) Interaction energies between the proteins and ZK60/AZ31 Mg alloys based on the MD simulations. The 1fid, 1ttg, and 3k65 correspond to the fibrinogen, fibronectin, and prothrombin, respectively. (Blue line represents the total interaction energy; orange line represents the vdW energy; yellow line represents the electrostatic interaction energy).

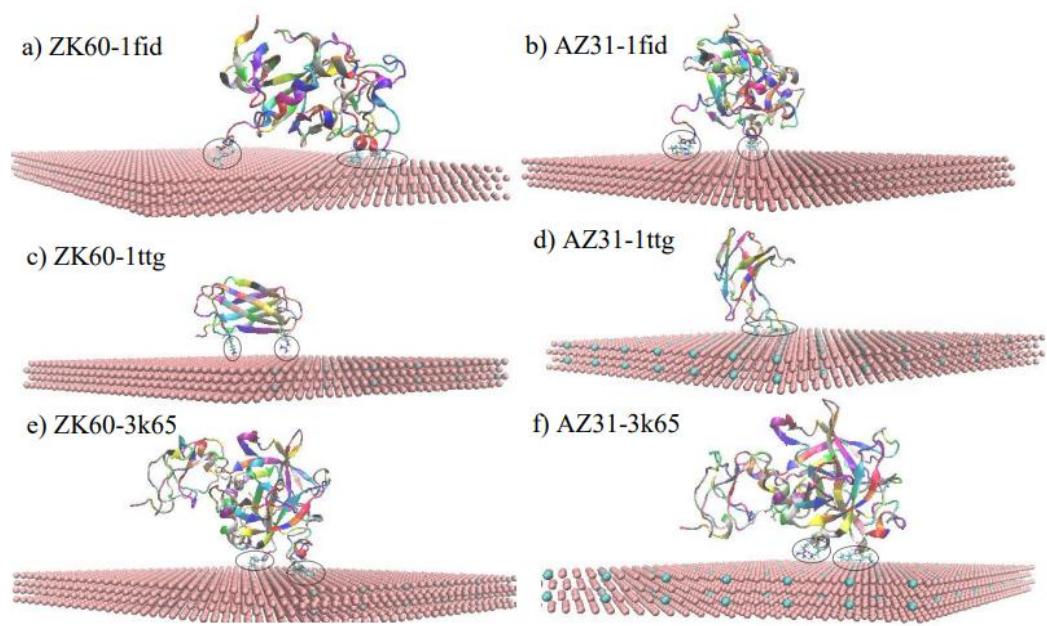

Fig. 9 a–f) Overall conformations of the three proteins (fibrinogen, fibronectin, and prothrombin) adsorbed on the surface of AZ60 and AZ31, the residues within 3.5 Å from the surface are represented by the Licorice pattern, respectively.